\documentstyle [12pt]{article}
\begin{document}
\normalsize
\centerline{\bf ON THE CALCULATION OF OHMIC LOSSES AT THE METALLIC}
\centerline{\bf SURFACE WITH SHARP EDGES} 
\normalsize
\vskip 5mm
\centerline{\it Alexander M. Dykhne}
\small
\centerline{TRINITI}
\centerline{142092 Troitsk, Moscow Region; 
e-mail: Sasha\_ dykhne@hotmail.com}
\vskip 3mm
\centerline{\it Inna M. Kaganova}
\small
\centerline{Institute for High Pressure Physics Russian Academy of 
Sciences}
\centerline{142092 Troitsk, Moscow Region; e-mail: 
u10740@dialup.podolsk.ru}
\normalsize
\begin{abstract}
We discuss the applicability of the perturbation theory in 
electrodynamic problems where the local Leontovich (the impedance) boundary 
conditions are used to calculate the ohmic losses at the metallic 
surface. As an example, we examine a periodic grating formed from semi-
infinite rectangular plates exposed to the s-polarized electromagnetic 
wave. Two different ways for calculation of the ohmic losses are 
presented: (i) the calculation of the reflection coefficient obtained 
with the aid of the perturbation theory (the impedance is the small 
parameter) and (ii) the direct calculation of the energy flux through 
the metallic surface, when to get the answer only the tangential 
magnetic field at the surface of a perfect conductor of the same 
geometry has to be known. The results (i) and (ii) differ noticeably.
The same difficulty is inherent in all the problems where the metallic 
surface has rectangular grooves. We show that the standard first order 
perturbation theory is not applicable since beginning from a number $n$ even the 
first corrections to the modal functions ${\phi}_{n}$ used to calculate 
the fields, are of the same order as the zero order modal functions (the 
impedance is equal to zero). Basing on the energy conservation law we 
show that the accurate value for the ohmic losses is obtained with the 
aid of the approach (ii). 
\end{abstract} 

\section {INTRODUCTION}

In the present publication we would like to fix the readers attention on 
an unexpected situation arising when the perturbation theory is used for 
calculation of the ohmic losses at the highly conducting metallic 
surfaces with sharp edges. As an example we examine the one-dimension 
metallic periodic grating formed from semi-infinite rectangular plates 
(see Fig.1) that is illuminated by s-polarized light.

All the calculations are performed in the framework of the local 
impedance boundary conditions (the Leontovich boundary conditions) 
applicability \cite{1,2}. In other words, we assume that 
$$
\delta \ll b \ll \lambda,  \eqno (1)
$$
where $\delta$ is the penetration depth of an electromagnetic wave into 
the metal, $b$ is the characteristic size of the surface contour (in our 
case, the period of the grating) and $\lambda = 2\pi c/\omega$ is the 
vacuum wavelength, $\omega$ is the frequency of the incident wave. 

This allows us to restrict ourselves to solution of the Maxwell 
equations in the region external with respect to a metal. At the 
metallic surface the boundary conditions for the fields are 
$$
{\bf E}_{t} = \hat \zeta [{\bf n},{\bf H}_{t}],    \eqno (2)
$$
${\bf E}_{t}$ and ${\bf H}_{t}$ are tangential components of electric 
and magnetic vectors at the metallic surface, $\bf n$ is the unit 
external normal vector to the surface. The two-dimensional tensor 
$\hat\zeta$ is the surface impedance tensor. The real and imaginary parts of
the impedance define the dissipated energy and the phase shift of the reflected
electromagnetic wave respectively (see, for example, \cite {1}). In what follows
we suppose that in Eqs.(2) the impedance $\hat\zeta$ is an ordinary
multiplying operator, and it is the same as the impedance of a perfectly flat 
surface of the same metal. 

Of course, the boundary conditions (2) are not rigorous ones. In general 
the relation between ${\bf E}_{t}$ and ${\bf H}_{t}$ is nonlocal. 
Moreover, when the metallic surface is not flat, even the local terms in 
the expression for the components of the surface impedance tensor have 
corrections depending on the shape of the metallic surface (see, for 
example, \cite{3,4}.) However, when the inequalities (1) are fulfilled, 
the leading term in the expression for $\hat \zeta$ is equal to the 
impedance of the perfectly flat metallic surface. Having in mind to 
calculate the leading term in the expression for the ohmic losses we 
restrict ourselves to these approximation only.

In what follows we assume that the metal is an isotropic one. Then the 
surface impedance tensor is ${\zeta}_{\alpha\beta} = 
\zeta{\delta}_{\alpha\beta}$. Under the conditions of normal skin effect 
\cite{1} $\zeta = \sqrt{\omega/8\pi\sigma}(1-i)$, where $\sigma$ is the 
conductivity of the metal. Under the conditions of anomalous skin effect 
the expression for the surface impedance is more complex (see, for 
example, \cite{5}). In our calculations we do not specify the character 
of skin effect. We only take into account that in the order of magnitude 
$|\zeta| \sim \delta/\lambda$. For good metals this ratio is extremely 
small.

Note, the local impedance boundary conditions (2) are valid up to the 
terms of the order of $|\zeta|$ only. Consequently, when calculating the 
fields the terms of the higher order (for example, of the order 
$b|\zeta|/\lambda$, ${|\zeta|}^{2}$ and so on) must be omitted as far as 
they exceed the accuracy allowed by the boundary conditions. 

Strictly speaking, inequalities (1) provide the possibility to use the 
local boundary conditions, Eqs.(2), when calculating the fields above 
the metal only if the surface of the metal is rather smooth. The 
presence of sharp edges makes this possibility questionable. Of course, 
in real life all the sharp edges are rounded ones. Undoubtedly, the 
boundary conditions (2) can be used only if the smoothing characteristic 
size is greater than $\delta$ too. However, it is rather difficult to 
take the smoothing regions into account accurately even for very simple 
surface profiles. Therefore rather often Eqs.(2) are used, assuming that 
a smoothing changes the results only slightly.

It must be noted that the grating depicted in Fig.1 is infinitely long 
in the $x_3$ direction. At first it seems that the boundary conditions 
(2) are not valid in this case. To justify ourselves we take into 
account that in the s-polarization state the fields decay exponentially 
inside the grooves when the distance from the throat of the groove 
increases. As a result the fields in the vicinity of infinitely deep 
grooves and the grooves of finite, but rather large depth $h$ ($h\ge ma$ 
and  $m$ is of the order of some units) are practically the same. For 
such grooves the smoothing of the contour allowing us to use the 
boundary conditions (2) can be done rather easily.

After all these preliminary remarks we are faced with a rigorously set 
mathematical problem: suppose that a plane wave ${\bf E}^{i} = {\bf 
E}_{0}{\rm e}^{-i(kx_3+\omega t)}$ ($k = \omega/c$) is normally incident 
upon the surface depicted in Fig.1. At the metallic surface we have the 
boundary conditions (2) that are valid up to the terms of the order 
$|\zeta|$. We impose an outgoing wave or exponentially decaying boundary 
conditions on the reflected fields. Our problem is to define the ohmic 
losses. In other words, we need to calculate the energy flux through the 
surface of the metal.

Let us introduce the time-averaged Pointing vector in the region outside 
the metal 
$$
\overline{\bf S} = (c/8\pi ){\rm Re}[{\bf E}{\bf H}^{*}] \eqno (3)
$$
that is the density of the time-averaged energy flux. According to the 
energy conservation law \cite{1}
$$
{\rm div}\overline{{\bf S}}=0. \eqno (4)
$$
The consequence of the last equation is the equality of the integrals
$$
\int_{S_{\infty}} \overline{\bf S} {\rm d}{\bf f} = 
\int_{S_m} (\overline{{\bf S}}{\bf n}){\rm d}f,  \eqno (5)
$$
where the integral in the left-hand side is taken over the surface 
infinitely distant from the metal. The integral in the right-hand side 
is taken over the surface of the metal. It defines the energy flux 
penetrating through the metallic surface. 

It is well known that infinitely far from the metallic surface ($x_3 \gg 
\lambda$) the only nonzero component of the Pointing vector is expressed 
in terms of the reflection coefficient $R$: ${\overline S}_{3} = 
-(c/8\pi){|{\bf E}_{0}|}^{2}(1-R)$. Consequently, when the reflection 
coefficient is known, we can calculate the ohmic losses with the aid of 
the left-hand side of Eq.(5).

On the other hand we can calculate the energy flux through the surface 
of the metal directly. With regard to the boundary conditions (2) at the 
metallic surface the normal component of the time-averaged Pointing 
vector is ${\overline S}_n = \frac{c}{8\pi}{\rm Re}\zeta 
{|{\bf H}_{t}|}^{2}$. The last expression can be simplified.

Since $|\zeta| \ll 1$, the magnetic vector ${\bf H}_{t}$ at the surface 
of a good metal is approximately the same as the magnetic vector 
${\bf H}_{t}^{(0)}$ at the surface of the perfect conductor (the 
conductivity is infinitely large and, consequently, the impedance $\zeta 
=0$) of the same geometry \cite{1}. Then within the allowed accuracy we 
must rewrite ${\overline{S}}_n$ as ${\overline{S}}_n = 
\frac{c}{8\pi}{\rm Re}\zeta {|{\bf H}_{t}^{(0)}|}^{2}$.

Let us introduce the energy $Q_d$ dissipating in the metal per unit 
length in the $x_2$ direction and per one period. According to the 
previous argumentation and making use of Eq.(5) we can write two 
equivalent expressions for $Q_d$
$$
{Q}_{d} = \frac{c}{8\pi}{\rm Re}\zeta \int_{L} 
{|{\bf H}_{t}^{(0)}|}^{2} {\rm d}l;  \eqno (6.a)
$$
the integral is taken over the contour of the surface relating to one 
period, and 
$$
{Q}_{d}  = 2b{|E_0|}^{2}\frac{c}{8\pi}[1-R]. \eqno (6.b)
$$

Note, when Eq.(6a) is used, to calculate $Q_d$ we need to know the 
magnetic vector ${\bf H}_{t}^{(0)}$ at the surface of the perfect 
conductor only. On the other hand, when we make use of Eq.(6.b) we need 
to know the reflection coefficient $R$. To calculate $R$ we need the 
solution of the complete electrodynamic problem in the region over the 
real metal. 

A lot of authors calculated the fields above perfectly conducting 
gratings (see, for example, \cite{6}). It can be done with the aid of 
different methods \cite{7}. For example, the fields inside the grooves 
are presented as series of modal functions that are the solutions of the 
Maxwell equations. The representation incorporates the boundary 
conditions for perfect conductors. Above the grooves the usual Rayleigh 
plane-wave representation is adopted. As a result a matrix equation for 
the modal amplitudes is obtained. This method makes use of the standard 
Fourier analysis. There are no doubts with respect to the results 
obtained with the aid of this method. The other way is based on the 
solution of the Wiener-Hopf equation. Both methods give the same result. 

Applying the local impedance boundary conditions, we can use the same 
methods when calculating the fields above highly conducting metallic 
gratings. However, to be sure that we do not fall outside the limits of 
the boundary conditions (2) applicability, when calculating we have to 
exploit the perturbation theory based on the inequalities (1). In what 
follows we show that in this problem the standard perturbation theory is 
not applicable. More accurately, we show that in our problem the 
dissipated energy $Q_d$ calculated with the aid of Eq.(6.a) differs from 
the result obtained with the aid of Eq.(6.b) after calculation of the reflection 
coefficient $R$ in the framework of the perturbation theory. This discrepancy 
leads to violation of the energy conservation low. 

We would like to stress once more, the field ${\bf H}_{t}^{(0)}$ 
entering Eq.(6.a) can be calculated with high accuracy. Then, in the 
framework of our approximation Eq.(6.a) provides the correct result for 
the ohmic losses. Next, as it is shown below, the method of calculation 
of the fields above the real metal (the impedance $\zeta \ne 0$) is much 
alike the one used when calculating the fields above the perfect 
grating. However, to calculate the difference $1-R$ up to the terms of 
the order of $\zeta$, we have to use the perturbation theory with 
respect to the small impedance $\zeta$. Consequently, the source of the 
obtained discrepancy can be the perturbation theory only.

We faced this problem when calculating the ohmic losses on the lamellar 
grating with finite depth of the grooves exposed to the s-polarized 
light \cite{8}. We found out that for rather deep grooves our result was 
approximately twenty percents less than the result obtained by 
L.A.Vainshtein, S.M.Zhurav and A.I.Sukov in \cite{9}. In the same frequency
region defined by Eqs.(1) in the framework of the impedance boundary conditions
(2) these authors examined the one-dimensional grating with infinitely deep
grooves and very thin plates ($(b-a)/b \ll 1$). They based their solution on
the application of the Wiener-Hopf equation and calculated the reflection
coefficient. Then taking into account that according to Eqs.(1) the parameter 
$|\zeta |/kb \ll 1$, they used the perturbation theory and calculated the 
difference $1-R$ entering Eq.(6.b) up to the terms of the order of $\zeta$.

In the s-polarization state there are no physical reasons of different results
for very deep and infinitely deep grooves. Therefore, trying out to verify our
approach, for the grating with infinitely deep grooves we reproduced the
perturbation theory calculation of \cite{9} with the aid of the modal function
method adopted in our calculations in \cite{8}. The results obtained with the 
aid of the perturbation theory were the same as in the work of Vainshtein and 
his co-authors. Then we repeated the calculation with the aid of Eq.(6.a).
Comparing the results we found the same twenty percents difference (see Fig.2).

The result of this analysis is presented in this publication. From our point of
view our calculations show that in this problem the application of the standard
perturbation theory provide a regular error when calculating the fields above
the grating. In what follows we show the place where the standard perturbation
theory failed. 

It was not the goal of this work to develop a regular perturbation theory
allowing us to calculate the fields above lamellar conducting gratings
correctly. The main results of our investigation can be formulated as follows:

\noindent i. For highly conducting lamellar gratings the correct result 
for ohmic losses $Q_d$ is given by Eq.(6.a).

\noindent ii. The coincidence of the result for $Q_d$ obtained with the 
aid of Eq.(6.a) and Eq.(6.b) can be used as a proof of correct 
calculation of the fields above the metallic surface.

The organization of this paper is as follows. In Section 2 we present 
the modal function approach allowing us to calculate the fields above 
the metallic grating. In Section 3 we describe the results obtained in 
the framework of the standard perturbation theory. In Section 4 we 
calculate the ohmic losses and discuss the reasons why the standard 
perturbation theory fails. Concluding remarks are given in Section 
5. In APPENDIX we show that for an arbitrary value of $kb < 1$, the 
formulae used when calculating the fields above the infinitely 
conducting grating provide the total reflection of the incident wave 
(the reflection coefficient $R=1$). 

\section {CALCULATION OF THE FIELDS}

Let the one-dimensional periodic metallic grating depicted in Fig.1 be 
exposed to the s-polarized normally incident electromagnetic wave ${\bf 
E}^{i} = (0,{E}_{0},0){\rm exp}[ - ik{x}_{3}]$. For simplicity we assume 
that on the flat parts $x_3 = 0$ of the surface the impedance is equal 
to zero. On the inner surfaces of the semi-infinite grooves the boundary 
conditions (2) are fulfilled.

The inequalities (1) provide us with three small parameters. Namely,
$$
|\zeta| \sim \frac{\delta}{\lambda} \ll 1, \quad kb \ll 1 \quad 
{\rm and} \quad \frac{\delta}{b} \sim  \frac{|\zeta|}{kb} \ll 1.
\eqno (7)
$$
No other restrictions are imposed on the ratio $a/b < 1$.

At first, we will not pay attention to inequalities (7). In what follows 
we obtain the solution of the problem without using the perturbation 
theory. We assume only that $kb < 1$.

Let the superscripts $+$ and $-$ denote the fields in the half-space 
$x_3 > 0$ and in the central groove respectively. With regard to the 
surface periodicity we adopt the representation 
$$
{E}^{+}({x}_{1},{x}_{3}) = {E}_{0}\left\{
{\rm e}^{-ikx_3} + \sum_{q = -\infty}^{\infty} {e}_{q}^{+}{\rm e}^{i(\pi 
q{x}_{1}/b + {\alpha}_{q}{x}_{3})}\right\},
\eqno (8)
$$ 
for the electric field above the metal. In Eq.(8) ${\alpha}_{0} = k$ and 
${\alpha}_{q} = i\sqrt{{(\pi q/b)}^{2}-k^2}$, if $q \ne 0$. Since on the 
plain parts $x_3 = 0$ of the surface the impedance is equal to zero, 
${E}_{2}^{+}({x}_{1},0) = 0$, when $a < |x_1| < b$.  

Inside the central groove ($|x_1| \le a$, $x_3 < 0$) we present 
${E}^{-}({x}_{1},{x}_{3})$ as a series of modal functions 
${\phi}_{n}(x_1,x_3)$ that are the solutions of the Maxwell equations 
incorporating the boundary conditions (2) on the vertical sides of the 
groove\footnote{An analogous method was repeatedly used when calculating  
the fields above infinitely conducting rectangular gratings (see, for 
example, \cite{6})}. Let ${B}_{n}$ be the coefficients of this series. 

Separating the variables we write the modal functions 
${\phi}_{n}(x_1,x_3)$ as
 $$
{\phi}_{n}(x_1,x_3) = {\psi}_{n}(x_3){\varphi}_{n}(x_1); \quad 
{\psi}_{n}(x_3) = {\rm exp}\left[\frac{x_3}{a}
\sqrt{{\gamma}_{n}^{2}-{(ka)}^{2}}\right]. \eqno (9.1)
$$
It is easy to see that the functions ${\varphi}_{n}(x_1)$ are given by 
different expressions for even and odd numbers $n$:
$$
{\varphi}_{2n}(x_1) = \frac{\sin ({\gamma}_{2n}x_1/a)}
{\cos {\gamma}_{2n}}; \quad  {\varphi}_{2n-1}(x_1) = 
-\frac{\cos ({\gamma}_{2n-1}x_1/a)}{\sin {\gamma}_{2n-1}} \eqno (9.2)
$$
with ${\gamma}_{2n}$ and ${\gamma}_{2n-1}$ being the solutions of 
dispersion equations
$$
\frac{\tan {\gamma}_{2n}}{{\gamma}_{2n}} = - \frac{i\zeta}{ka}; \quad 
\frac{\cot {\gamma}_{2n-1}}{{\gamma}_{2n-1}} =  \frac{i\zeta}{ka}.  
\eqno (9.3)
$$

The set of modal functions ${\phi}_{n}(x_1,x_3)$ has a lot of 
interesting properties. We do not discuss them here, but 
refer the reader to the review work of Yu.I.Lyubarskii \cite{10}.   
Note, the functions ${\varphi}_{n}(x_1)$ are not orthogonal ones 
since their scalar product $({\varphi}_{m},{\varphi}_{n}) = 
(1/2a)\int_{-a}^{a}{\rm d}x_1 {\varphi}_{m}^{*}(x_1){\varphi}_{n}(x_1)$ 
(the star means the complex conjugate) does not equal to zero when $m \ne n$.
However, the functions ${\varphi}_{n}(x_1)$ with different subscripts are
orthogonal:  
$$
\frac{1}{2b}\int_{-a}^{a}{\rm d}x_1 
{\varphi}_{m}(x_1){\varphi}_{n}(x_1) = {I}_{mm}{\delta}_{mn}, \eqno 
(10.1)
$$ 
where
$$
{I}_{2m,2m} = \frac{a}{2b}\left\{\frac{1}{\cos^2 {\gamma}_{2m}} +
\frac{i\zeta}{ka}\right\}; \quad
{I}_{2m-1,2m-1} =\frac{a}{2b}\left\{\frac{1}{\sin^2 {\gamma}_{2m-1}} +
\frac{i\zeta}{ka}\right\}.  \eqno (10.2)
$$
We used this property of the functions ${\varphi}_{n}(x_1)$ when 
deriving the matrix equation for the modal coefficients (see below 
Eq.(13)).

Now with regard to Eqs.(9) we write the field ${E}^{-}({x}_{1},{x}_{3})$ 
($|x_1|<a,\, x_3<0$) as
$$
{E}_{2}^{-}({x}_{1},{x}_{3}) ={E}_{0}\left\{\sum_{n=0}^{\infty} 
{B}_{2n}{\psi}_{2n}(x_3){\varphi}_{2n}(x_1) +
\sum_{n=1}^{\infty} {B}_{2n-1}{\psi}_{2n-1}(x_3)
{\varphi}_{2n-1}(x_1)\right\}. \eqno (11)
$$

Applying the boundary conditions on the plateaus $x_3=0$ and the 
continuity conditions across the central slit, we obtain two linked 
matrix equations allowing us to determine both the amplitudes $e_q^+$ 
and the modal amplitudes $B_n$. In place of the amplitudes $B_n$ it is 
convenient to introduce the coefficients $X_n$,
$$
i(kb)X_n = \frac{2{\gamma}_{n}}{\pi}B_n.  \eqno (12)
$$

We eliminate the amplitudes ${e}_{q}^{(+)}$ from the set of matrix 
equations to obtain the matrix equation for the coefficients $X_n$.
There are two separate matrix equations for the coefficients $X_{2n}$ 
and $X_{2n-1}$. The even coefficients $X_{2n}$ satisfy a homogeneous matrix 
equation. Consequently, all the coefficients $X_{2n}=0$. For the 
odd coefficients $X_{2n-1}$ our calculations results in the following 
matrix equation:
$$
\sum_{m=1}^{\infty}{X}_{2m-1}{\Delta}_{2m-1,2l-1} + 
{\left(\frac{\pi b}{2a}\right)}^{2}
\left(\frac{\pi}{2{\gamma}_{2l-1}}\right)
\sqrt{1-{(ka/{\gamma}_{2l-1})}^{2}}{I}_{2l-1,2l-1}{X}_{2l-1} = 
2{W}_{2l-1}. \eqno (13)
$$
If we set $\mu = 2a/b$, the matrix ${\Delta}_{2m-1,2l-1}$ in Eq.(13) is
$$
{\Delta}_{2m-1,2l-1}=\sum_{q=1}^{\infty}q\mu \sqrt{1- {(kb/\pi q)}^{2}}
{D}_{2m-1,2l-1}(q) - i\frac{\mu}{\pi}(kb)
{W}_{2l-1}{W}_{2m-1}; \eqno (14.1)
$$
$$
{D}_{2m-1,2l-1}(q) = 2{\left(\frac{\pi}{\mu}\right)}^{2}
{C}_{2l-1,q}{C}_{2m-1,q}; \quad 
{W}_{2l-1} ={\left(\frac{\pi}{2{\gamma}_{2l-1}}\right)}^{2},
 \eqno (14.2)
$$
${C}_{2l-1,q}$ stands for the integral
$$
\left(\frac{2{\gamma}_{2l-1}}{\pi}\right){C}_{2l-1,q} = \frac{1}{2b}
\int_{-a}^{a}{\rm d}x_1 {\rm e}^{-i\pi qx_1/b} {\varphi}_{2l-1}(x_1). 
 \eqno (14.3)
$$
The explicit form of ${C}_{2l-1,q}$ is
$$
{C}_{2l-1,q} = \left(\frac{\mu\pi}{4}\right)
\frac{\cos {\omega}_{q}-{\omega}_{q}\sin [{\omega}_{q}
\cot {\gamma}_{2l-1}/{\gamma}_{2l-1}]}
{{\omega}_{q}^{2} - {\gamma}^{2}_{2l-1}}, \quad 
{\omega}_{q} = \frac{\pi q\mu}{2},  \eqno (14.4)
$$
and ${I}_{2l-1,2l-1}$ is defined by Eq.(10).

We can calculate the coefficients $X_{2n-1}$ solving numerically the 
infinite set of equations (13). To calculate the reflection coefficient 
$R$ we take into account that since $kb < 1$, there is only one 
reflected wave. In accordance with Eq.(8) its dimensionless amplitude is 
$e_0^+$ and $e_q^+$ with $q \ne 0$ represent evanescent waves. Omitting 
intermediate calculations, we present the expression for $e_0^+$ in 
terms of the coefficients $X_{2l-1}$:
$$
e_0^+ = -1 -i(kb)\frac{\mu}{\pi}\sum_{l=1}^{\infty}X_{2l-1}W_{2l-1}.
\eqno (15)
$$
Then $R = {|e_0^+|}^{2}$.

It seems, that in this way we solve the problem and can calculate the 
ohmic losses substituting the found reflection coefficient into 
Eq.(6.b). However, as it was aforementioned, the local impedance 
boundary conditions (2) are true up to the terms of the order of $\zeta$ 
only. Consequently, to obtain a reliable answer for the ohmic losses we 
need to extract the terms of the order of $\zeta$ from the expression 
for $1-R$. And this is just the place where the perturbation theory has 
to be used.

\section {THE RESULT OF THE FIRST ORDER PERTURBATION THEORY}

The zero term of our perturbation theory corresponds to $\zeta = 0$. We 
use the superscript $(0)$ to denote that $\zeta = 0$. Then in accordance 
with Eqs.(9) we have
$$
{\gamma}_{2n-1}^{(0)} = \frac{\pi (2n-1)}{2}; \quad
{\varphi}_{2n-1}^{(0)}(x_1) = {(-1)}^{n}\cos (\pi {(2n-1)}x_1/2a).\eqno 
(16)
$$
The functions ${\varphi}_{2n-1}^{(0)}(x_1)$ are cosines, so they can be 
used as basic functions for the standard Fourier analysis.
 
Next, from Eq.(9.3) it follows that the eigenvalues ${\gamma}_{2l-1}$ 
depend not on $\zeta$ itself, but on the parameter $\varepsilon  = \zeta 
/(ka)$. Since $ka \ll 1$, it is evident that $|\zeta |\ll 
|\varepsilon |\ll 1$. With regard to Eq.(9.3) we obtain that up to the 
terms of the order of $\varepsilon$
$$
{\gamma}_{2l-1} = {\gamma}_{2l-1}^{(0)}(1-i\varepsilon). \eqno (17)
$$
Moreover, the impedance $\zeta$ enters the matrix equation (13) only 
through the dependence on ${\gamma}_{2l-1}$. Thus, from Eq.(13) it 
follows that the coefficients $X_{2n-1}$ depend not on all the three 
small parameters listed in Eq.(7) separately, but on the two small 
parameters only. They are $kb$ and $\varepsilon$. When $\varepsilon = 0$ 
the coefficients ${X}_{2n-1}^{(0)}(kb)$ define the fields above the 
infinitely conducting grating. With regard to Eqs.(13) and Eqs.(14) it 
is easy to see that the matrix equation for ${X}_{2n-1}^{(0)}(kb)$ is
$$
\sum_{m=1}^{\infty}{X}_{2m-1}^{(0)}{\Delta}_{2m-1,2l-1}^{(0)} + 
\frac{1}{\mu}{\left(\frac{\pi}{2}\right)}^{2}\sqrt{1-{(kb\mu /\pi 
(2l-1))}^{2}}\frac{{X}_{2l-1}^{(0)}}{2l-1} = \frac{2}{{(2l-1)}^{2}}, 
\eqno 
(18.1)
$$
where
$$
{\Delta}_{2m-1,2l-1}^{(0)}(kb) = \sum_{q=1}^{\infty}q\mu 
\sqrt{1- {(kb/\pi q)}^{2}}{D}_{2m-1,2l-1}^{(0)}(q) - 
i\frac{\mu}{\pi}(kb)\frac{1}{{(2m-1)}^{2}{(2l-1)}^{2}}, \eqno (18.2)
$$
and 
$$
{D}_{2m-1,2l-1}^{(0)}(q) = \frac{1+\cos 2{\omega}_{q}}{d_m(q)d_l(q)}; 
\quad 
d_m(q) = {(q\mu)}^{2}-{(2m-1)}^{2}, \eqno (18.3)
$$
${\omega}_{q}$ is defined in Eq.(14.4).

Suppose that we know the solution of Eq.(18.1). To take into account the 
finite conductivity of the metal, we need to make use of the 
perturbation theory when solving the matrix equation (13) and calculate 
the first corrections to ${X}_{2n-1}^{(0)}(kb)$ in the small parameter 
$\varepsilon $. 

Note, according to Eq.(11) and the definition of the coefficients 
$X_{2n-1}$, Eq.(12), the same corrections provide the electric field 
${E}^{-}$ up to the terms of the order of $\zeta$. Since the starting 
point of these calculations, the local impedance boundary conditions 
(2), do not allow us to calculate the fields up to the terms higher than 
$\zeta$, when calculating the coefficients ${X}_{2n-1}$ the terms of the 
order higher than $\varepsilon$ (for example, of the order of $\zeta$) 
must be omitted within the accuracy of the local impedance boundary 
conditions applicability. 

With the aid of Eq.(17) we expand all the coefficients in Eq.(13) up to 
the terms of the order of $\varepsilon$. Then with regard to Eqs.(18) 
after tedious, but straightforward algebra we obtain that up to the 
terms of the order of $\varepsilon$ the solution of Eq.(13) has the form
$$
{X}_{2n-1}(kb;\varepsilon) = {X}_{2n-1}^{(0)}(kb) + 
2i\varepsilon ({X}_{2n-1}^{(0)}(0) - {V}_{2n-1}), \eqno (19)
$$
where the coefficients ${V}_{2n-1}$ satisfy the matrix equation 
$$
\sum_{m=1}^{\infty}{V}_{2m-1}{\Delta}_{2m-1,2l-1}^{(0)}(0) + 
\frac{1}{\mu}{\left(\frac{\pi}{2}\right)}^{2}\frac{{V}_{2l-1}}{2l-1} = 
\sum_{m=1}^{\infty}{X}_{2m-1}^{(0)}(0){S}_{2m-1,2l-1}; \eqno (20.1)
$$ the matrix ${\Delta}_{2m-1,2l-1}^{(0)}(0)$ is defined by Eq.(18.3) when 
$kb=0$, the matrix ${S}_{2m-1,2l-1}$ is
$$
{S}_{2m-1,2l-1} = \sum_{q=1}^{\infty}(q\mu ){T}_{2m-1,2l-1}(q) + 
\frac{1}{\mu}{\left(\frac{\pi}{2}\right)}^{2}\frac{1}{2l-
1}{\delta}_{m,l}, 
\eqno (20.2)
$$ 
and
$$
{T}_{2m-1,2l-1}(q) = -\frac{1}{d_m(q)d_l(q)}\left[ (1+\cos 
2{\omega}_{q})
\left( \frac{{(2m-1)}^{2}}{d_m} + \frac{{(2l-1)}^{2}}{d_l} \right)
+ {\omega}_{q}\sin 2{\omega}_{q} \right].  \eqno (20.3)
$$
Again ${\omega}_{q} = \pi q\mu /2$ and $d_m$ is defined in Eq.(18.3).

If in consecutive order we solve Eqs.(18) and (20), we obtain the 
coefficients $X_{2n-1}(kb,\varepsilon)$ up to the terms linear in the 
parameter $\varepsilon$. 

\section {CALCULATION OF OHMIC LOSSES}

At first, let us calculate the dissipated energy $Q_d$ with the aid of 
Eq.(6.a). We mark the result of this calculation with the superscript 
$(sur)$. We remind that we set $\zeta = 0$ at the flat parts $x_3=0$ of 
our surface. Then in Eq.(6.a) the integration is performed over the 
vertical facets of the central groove, where the tangential magnetic 
vector is ${H}_{3}^{(0)}(\pm a,x_3)$ ($x_3 < 0$). Now we can rewrite 
Eq.(6.a) as
$$
{Q}_{d}^{(sur)} = \frac{c}{4\pi}{\rm Re}\zeta \int_{-\infty}^{0} 
{|{H}_{3}^{(0)}(a,x_3)|}^{2} {\rm d}x_3;  \eqno (21)
$$
We take into account that since the right-hand side of Eq.(21) is 
proportional to $\zeta$, within the accuracy of our approach the field 
${H}_{3}^{(0)}(a,x_3)$ in the integrand has to be known only up to the 
terms independent of the small parameter $kb$. Then with regard to the 
Maxwell equations and equations (11) and (12) (we remind that all the 
coefficients $B_{2n}=0$) we have 
$$
{H}_{3}^{(0)}(a,x_3){|}_{kb=0} = -\frac{\pi}{\mu}{E}_{0}
\sum_{l=1}^{\infty}{X}_{2l-1}^{(0)}(0){\rm exp}[\pi(2l-1)x_3/2a],  
\eqno (22)
$$
and, consequently,
$$
{Q}_{d}^{(sur)}= cb{|E_0|}^{2}
\frac{{\rm Re}\zeta}{8\mu}\sum_{l,m=1}^{\infty}
\frac{{X}_{2l-1}^{(0)}(0) {X}_{2m-1}^{(0)}(0)}{l+m-1}.  \eqno (23)
$$

Now let us calculate the energy $Q_d$ with the aid of Eq.(6.b). We mark 
this result with the superscript $(\infty)$. Since the reflection 
coefficient $R = {|e_0^+|}^{2}$, to calculate $Q_d^{(\infty)}$ we use 
Eq.(15). Then 
$$ 
Q_d^{(\infty)} = -\frac{cb{|E_0|}^{2}}{4\pi}Q; 
$$
$$
Q = i(kb)\frac{\mu}{\pi}\sum_{l=1}^{\infty}({X}_{2l-1}W_{2l-1}-
{X}_{2l-1}^{*}W_{2l-1}^{*}) +
{(kb)}^{2}{\left(\frac{\mu}{\pi}\right)}^{2}\sum_{l,m=1}^{\infty}
X_{2l-1}^{*}X_{2m-1}W_{2l-1}^{*}W_{2m-1}.  \eqno (24)
$$ 

When calculating $Q_d^{(\infty)}$ it is necessary to have in mind, that 
the dissipated energy has to be equal to zero for a perfectly conducting 
surface. In other words, when $\zeta = 0$ and, consequently, $X_{2l-1} = 
X_{2l-1}^{(0)}(kb)$, $Q_d^{(\infty)}$ vanishes for an arbitrary value of
$kb<1$. In APPENDIX we show that this equality is fulfilled for 
$X_{2l-1}^{(0)}(kb)$ that are solutions of Eq.(18.1). 

As a result, substituting the expression (19) in Eq.(24) we obtain 
$Q_d^{(\infty)}$ in terms of the coefficients ${X}_{2l-1}^{(0)}(0)$ and 
${V}_{2l-1}$.
$$
Q_d^{(\infty)} = \frac{2cb{|E_0|}^{2}}{{\pi}^2}{\rm Re}\zeta
\sum_{l=1}^{\infty}\frac{2{X}_{2l-1}^{(0)}(0)-{V}_{2l-1}}{{(2l-1)}^{2}}. 
 \eqno (25)
$$
This equation is valid up to the terms of the order of $\zeta$

Our next step was numerical calculation of the coefficients\footnote{The 
computational procedure used in the present work when calculating the 
coefficients ${X}_{2l-1}^{(0)}(0)$ is just the same as the one used in 
\cite{8}. So we refer the readers to the work \cite{8}.} ${X}_{2l-
1}^{(0)}(0)$. With this result in hand we evaluated the right-hand side 
of the matrix equation (20.1) and calculated the coefficients $V_{2l-
1}$. Finally, we calculated numerically $Q_d^{(sur)}$ and  
$Q_d^{(\infty)}$ with the aid of Eq.(23) and Eq.(25) respectively.

In Fig.2 we plot the dimensionless values 
$q_i = \pi Q_d^{(\infty)}/cb{|E_0|}^{2}{\rm Re}\zeta$ and $q_s = \pi 
Q_d^{(sur)}/cb{|E_0|}^{2}{\rm Re}\zeta $ 
as functions of the ratio $a/b$. We see that for all $a/b <1$ the 
results for $Q_d^{(\infty)}$ and $Q_d^{(sur)}$ differ noticeably. 

In Fig.2 we also display the results for the ohmic losses obtained by 
L.A.Vainshtein and his co-authors in \cite{9}. As it was mentioned in 
INTRODUCTION, they examined the same grating formed from thin plates 
($(b-a)/b \ll 1$). In \cite{9} the way of calculation was much alike the 
one used in the present work when calculating $Q_d^{(\infty)}$, however 
some other modal functions were used, and the problem was reduced to 
solution of the Wiener-Hopf equation. Supposing that the impedance of 
the flat parts $x_3=0$ of the metallic surface was equal to zero, the 
authors of \cite{9} obtained an elegant analytic formulae for the 
leading term of the ohmic losses 
$$
Q_d^{(V)} = \frac{cb{|E_0|}^{2}}{{\pi}^{2}}{\rm Re}\zeta 
\ln \left(\frac{2a}{b-a}\right),\quad (b-a)/b \ll 1. \eqno (26)
$$
(The superscript $(V)$ indicates that this is the result of \cite{9}.) 

We see, that our results for $Q_d^{(\infty)}$ are in good agreement with 
the values $Q_d^{(V)}$. This confirms our calculations made in the 
framework of the perturbation theory. However, neither $Q_d^{(V)}$, nor 
$Q_d^{(\infty)}$ coincide with $Q_d^{(sur)}$ that is the true energy 
flux penetrating through metallic surface. Consequently, we conclude 
that the obtained result for $Q_d^{(\infty)}$ (the same as $Q_d^{(V)}$) 
is improper, since it leads to violation of the energy conservation law. 

\subsection{Inefficiency of the standard perturbation theory} 

We think that the source of this discrepancy is application of 
the standard perturbation theory. The point is that if $\zeta 
\ne 0$ the set of modal functions ${\phi}_{2n-1}(x_1,x_3)$ (see 
Eqs.(9)), used when calculating the reflection coefficient, being 
complete and minimal still does not form a stable basis. As a result expansions 
in this set can be very sensitive to perturbations (see, for example, 
\cite{10}).

To show this, let us note that according to Eqs.(16) the zero order eigenvalues 
${\gamma}_{2n-1}^{(0)}$ are equidistant numbers: ${\gamma}_{2n+1}^{(0)} - 
{\gamma}_{2n-1}^{(0)} = \pi$. However, the distance between ${\gamma}_{2n-1}$ 
defined by Eq.(17) and ${\gamma}_{2n-1}^{(0)}$ increases linearly when $n$ 
increases. If, for example, $\zeta = {\zeta}_{0}(1-i)$ (that is true for 
homogeneous solids under the conditions of normal skin effect), we have
$$
\frac{|{\gamma}_{2n-1} - {\gamma}_{2n-1}^{(0)}|}{\beta} = 
\frac{\pi}{\sqrt{2}}(2n-1), \quad \beta = {\zeta}_{0}/ka.  \eqno (27)
$$

Now let us calculate the eigenfunctions ${\varphi}_{2n-1}(x_1)$ 
substituting ${\gamma}_{2n-1}$ from Eq.(17) in Eq.(9.2). The result of 
this calculation is
$$
{\varphi}_{2n-1}(x_1) = {(-1)}^{n}
\frac{\cosh \beta{\gamma}_{2n-1}^{(0)}\frac{x_1}{a}
\cos (1+\beta ){\gamma}_{2n-1}^{(0)}\frac{x_1}{a} -
i \sinh \beta{\gamma}_{2n-1}^{(0)}\frac{x_1}{a}
\sin (1+\beta ){\gamma}_{2n-1}^{(0)}\frac{x_1}{a}}{\cos \beta {\gamma}_{2n-
1}^{(0)} \cosh \beta {\gamma}_{2n-1}^{(0)}
-i\sin \beta {\gamma}_{2n-1}^{(0)} \sinh \beta {\gamma}_{2n-1}^{(0)}}.  
\eqno (28)
$$
We see, that ${\varphi}_{2n-1}(x_1)$ is a complex function, whose real 
and imaginary parts are of the same order. The difference between 
${\varphi}_{2n-1}(x_1)$ and ${\varphi}_{2n-1}^{(0)}(x_1)$ is small only if $n 
\ll 1/\pi \beta$. When $n \gg 1/\pi \beta$ the eigenfunctions ${\varphi}_{2n-
1}(x_1)$ do not resemble the non-perturbed eigenfunctions ${\varphi}_{2n-
1}^{(0)}(x_1)$ at all. In Fig.3 for $x_1=0$ we plot the ratios ${\rm 
Re}{\varphi}_{2n-1}(0)/ {\varphi}_{2n-1}^{(0)}(0)$ and ${\rm Im}{\varphi}_{2n-
1}(0)/ {\varphi}_{2n-1}^{(0)}(0)$ for two values of $\beta$: $\beta = 0.05$ and
$\beta = 0.01$.

We must also take into account that in the s-polarization state the 
strength of the magnetic field in the immediate vicinity of a 
rectangular two-dimensional wedge increases significantly (see, for 
example, \cite{7}). When the fields are represented by the expressions 
of the type (11), to describe the fields near the corner points correctly we 
need to provide a special asymptotic behavior of the coefficients $X_{2n-1}$ 
when $n \to \infty$ (see \cite{7}). For $\zeta = 0$ the coefficients $X_{2n-
1}^{0} \sim 1/{(2n-1)}^{2/3}$ when $n \to \infty$ (see \cite{8})

With regard to all the aforementioned arguments we take for granted that 
in the s-polarization state the perturbation theory presented in Section 3 is 
not applicable when the local impedance boundary conditions are used to 
calculate the fields above one-dimensional periodic rectangular metallic 
gratings.

\section{CONCLUSIONS}

In this section we summarize the results of our analysis. We would like 
to remind that here we examined the case of the s-polarization only. We 
put aside the p-polarization state. In the end of this section we'll add 
some remarks relating to the p-polarization state.

From our point of view, the main result of this publication is the demonstration 
that the standard perturbation theory can fail if in the framework of the local 
impedance boundary conditions one try to use it when calculating the 
electromagnetic fields above periodic metallic gratings. We examined the case of 
rectangular infinitely deep grooves. However, it is quite possible that the same 
problems arise when the shape of the grooves is, for example, a triangular one, 
or, more general, when the surface profile has sharp edges.

Our calculations showed that for rough surfaces the results for the 
reflection coefficients obtained with the aid of the perturbation theory 
(the impedance $\zeta$ is a small parameter) are not always reliable. On 
the other hand, there are no doubts that the dissipated energy 
calculated with the aid of Eq.(6.a) defines the ohmic losses accurately.
Thus, we can define the reflection coefficient comparing equations (6.a) 
and (6.b). Then
$$
R = 1 - \frac{{\rm Re}\zeta}{2b{|E_0|}^{2}} 
\int_{L}{|{\bf H}_{t}^{(0)}|}^{2} {\rm d}l, \eqno (29)
$$
where the tangential magnetic field ${\bf H}_{t}^{(0)}$ is calculated up 
to the terms independent of $kb$. Let us remind that our results were 
obtained in the frequency region consistent with the inequalities (1).
 
On the other hand, this approach allows us to calculate the reflection 
coefficient only. The fields in the close vicinity of a rough metallic 
surface (at the distances $d$ that are of the order of the period of the 
surface structure where the evanescent waves are significant) cannot be 
calculated, if we know the expression for ${\bf H}_{t}^{(0)}$ only. To 
calculate this field one or another version of the perturbation theory 
must be used. 

We suggest to use Eq.(29) as a verification that the applied 
perturbation theory is accurate.

In some sense the calculation of the ohmic losses in the p-polarization 
state is much more simpl. The point is that if inequalities (1) are 
fulfilled the magnetic field penetrates into the grooves on the surface. 
In this case in the main approximation the magnetic field in the 
vicinity of the grooves is the same as near a flat surface of the 
perfect conductor. Thus the entire inner surface of the grooves absorbs 
the electromagnetic wave. In \cite{8} for an arbitrary one-dimensional periodic 
surface this argumentation allowed us to obtain rather simple expression for the 
leading terms of the effective surface impedance ${\zeta}_{ef}^{p}$ in the p-
polarization state:
$$
{\zeta}_{ef}^{p}={\rm Re}\zeta \frac{L}{2b} - i\frac{kS}{2b}. \eqno (30)
$$
In Eq.(30) $2b$ is the period of the surface structure, $L$ is the 
length of the surface profile per one period and $S$ is the area of the 
groove in the plane $x_1x_3$. If the effective impedance is known, there 
are no problems with calculation of the ohmic losses:
$$
Q_d^{(p)} = \frac{cb}{4\pi}{|H_0|}^{2}{\rm Re}{\zeta}_{ef}^{p}.
\eqno (31)
$$
However, some difficulties can arise when the perturbation theory is 
applied to calculate the evanescent waves near rough surface in the p-
polarization state. In the present publication we do not examine this 
question.

{\bf ACKNOWLEDGEMENTS}

The authors are grateful to prof. M.I.Kaganov, dr. T.A. Leskova, dr. 
L.B.Litinskii and prof. Yu.I.Lyubarskii for helpful discussions. The 
work of IMK was supported by RBRF grant 99-02-16533.

\section{APPENDIX}

Let us demonstrate that the solution $X_{2l-1}^{(0)}(kb)$ ($l = 1,2:$) 
of the matrix Eq.(18) guarantees the fulfillment of the energy 
conservation law for an arbitrary value of $kb < 1$. The equation (18) 
corresponds to the infinitely conducting grating ($\zeta =0$), and, 
consequently, the reflection coefficient has to be equal to one. With 
regard to our notations this means that substituting $X_{2l-
1}^{(0)}(kb)$ in Eq.(15) we have to verify that $1-R = 1-{|e_0^+|}^{2} = 
0$. When $\zeta = 0$, the explicit form of $1-R$ is
$$
1-R = -i(kb)\frac{\mu}{\pi}\sum_{l=1}^{\infty}\frac{(X_{2l-1}^{(0)} -
{( X_{2l-1}^{(0)})}^{*})}{{(2l-1)}^{2}}
-
{(kb)}^{2}{\left(\frac{\mu}{\pi}\right)}^{2}
\sum_{l,m=1}^{\infty}\frac{{( X_{2l-1}^{(0)})}^{*} X_{2m-1}^{(0)}}
{{(2l-1)}^{2}{(2m-1)}^{2}}\eqno (A.1)
$$
(compare with Eq.(24); $\mu = 2a/b$). When writing Eq.(A.1) we take into 
account that  $W_{2l-1}^{(0)} = 1/{(2l-1)}^{2}$ (see Eq.(14.2)).

We seek ${X}_{2m-1}^{(0)}(kb)$ in the form of a power series
$$
{X}_{2m-1}^{(0)} = \sum_{q=0}^{\infty}{Y}_{2m-1}^{(q)}{(kb)}^{q}.  \eqno 
(A.2)
$$
Then it is convenient to rewrite Eq.(18.1) in the form
$$
\sum_{r=1}^{\infty}{\Phi}_{2r-1,2m-1}{X}_{2r-1}^{(0)} = 
\frac{1}{(2m-1)}\left\{2 + i(kb)\frac{\mu}{\pi}
\sum_{r=1}^{\infty}\frac{{X}_{2r-1}^{(0)}}{{(2r-1)}^{2}}\right\}.  \eqno 
(A.3a)
$$
The elements of the matrix ${\Phi}_{2r-1,2m-1}$ depend on ${(kb)}^{2}$ 
only,
$$
{\Phi}_{2r-1,2m-1} = (2m-1)\sum_{q=1}^{\infty}q\mu \sqrt{1-{(kb/\pi 
q)}^{2}} 
{D}_{2r-1,2m-1}^{(0)} +\frac{1}{\mu}{\left(\frac{\pi}{2}\right)}^{2}
\sqrt{1-{\left(\frac{\mu kb}{\pi (2r-1)}\right)}^{2}}{\delta}_{rm},  
\eqno (A.3b)
$$
where ${D}_{2r-1,2m-1}^{(0)}$ is defined by Eq.(18.3). 

Let us present ${\Phi}_{2r-1,2m-1}$ as a series expansion:
$$
{\Phi}_{2r-1,2m-1} =\sum_{p=0}^{\infty}{\Phi}_{2r-1,2m-
1}^{(2p)}{(kb)}^{2p},
\eqno (A.4)
$$
When we substitute Eq.(A.2) in Eq.(A.3a) and set equal the terms, 
corresponding to the same powers of $(kb)$, we obtain a set of equations 
relating the coefficients ${Y}_{2m-1}^{(q)}$ with different values of 
the superscripts. Let us examine the obtained equations in the 
consecutive order. For $q=0$ (the term independent of $kb$) we have
$$
\sum_{r=1}^{\infty}{\Phi}_{2r-1,2m-1}^{(0)}{Y}_{2r-1}^{(0)} = 
\frac{2}{2m-1}. 
\eqno (A.5a)
$$
Evidently, ${Y}_{2r-1}^{(0)} = {X}_{2r-1}^{(0)}(0)$. Note, the 
coefficients ${Y}_{2r-1}^{(0)}$ being solutions of the linear set of 
equations with real parameters, are real numbers.

Next, for $q>1$ we have
$$
\sum_{p=0}^{s_q}\sum_{r=1}^{\infty}{\Phi}_{2r-1,2m-1}^{(2p)}
{Y}_{2r-1}^{(q-2p)} = \frac{2}{(2m-1)}{T}_{q-1}, \quad 
{T}_{q} = i\frac{\mu}{2\pi}\sum_{s=1}^{\infty}
\frac{{Y}_{2s-1}^{(q)}}{{(2s-1)}^{2}}.
 \eqno (A.5b)
$$
The first sum in the left-hand side of Eq.(A.5b) is taken over $p \le 
s_q$, where $s_q = s-1$, if $q=2s-1$, and $s_q = s$, if $q=2s$. 

The structure of the right-hand sides of Eqs.(A.5b) suggests to seek the 
coefficients ${Y}_{2s-1}^{(q)}$ ($q>1$) in the form
$$
{Y}_{2s-1}^{(q)}={Y}_{2s-1}^{(0)}{T}_{q-1}+{Z}_{2s-1}^{(q)}. \eqno (A.6)
$$
Then for $q=1$ we have
$$
{Z}_{2s-1}^{(1)} = 0; \quad  {Y}_{2s-1}^{(1)} = {Y}_{2s-1}^{(0)}{T}_{0}. 
\eqno (A.7a)
$$
We see that the coefficients ${Y}_{2s-1}^{(1)}$ are imaginary numbers. 
At the next step ($q=2$) the equation for ${Z}_{2s-1}^{(2)}$ is
$$
\sum_{r=1}^{\infty}{\Phi}_{2r-1,2m-1}^{(0)}{Z}_{2r-1}^{(2)} + 
\sum_{r=1}^{\infty}{\Phi}_{2r-1,2m-1}^{(2)}{Y}_{2r-1}^{(0)} = 0. \eqno 
(A.7b)
$$
With regard to Eq.(A.6) and Eqs.(A.7) it is evident that both ${Z}_{2r-
1}^{(2)}$ and ${Y}_{2r-1}^{(2)}$ are real numbers. When continuing this 
procedure, we find out that all the coefficients ${Y}_{2r-1}^{(2q-1)}$ 
are imaginary numbers and all the coefficients ${Y}_{2r-1}^{(2q)}$ are 
real numbers. On the contrary, $T_q$ is a real number when $q$ is odd, 
and $T_q$ is an imaginary number when $q$ is even.

In the consecutive order we wrote down equations for ${Z}_{2s-1}^{(q)}$. 
Examining these equations we obtained formulae relating ${Z}_{2s-
1}^{(q)}$ with the other coefficients ${Z}_{2s-1}^{(r)}$ with $r\le q$. 
At the last step we eliminated the functions ${Z}_{2s-1}^{(r)}$ with the 
aid of Eq.(A.6).

Here, omitting the intermediate calculations, we present only the 
recurrence formula for the coefficients ${Y}_{2r-1}^{(2q-1)}$ obtained 
with the aid of the aforementioned procedure:
$$
{Y}_{2r-1}^{(2q-1)} = i\frac{\mu}{2\pi}\sum_{s=1}^{\infty}\frac{1}{{(2s-
1)}^{2}}
\sum_{p=0}^{2(q-1)}{Y}_{2s-1}^{(p)}{({Y}_{2r-1}^{(2(q-1)-p)})}^{*}.  
\eqno (A.8)
$$
To obtain Eq.(A.8) we used the definition of the sum $T_q$ (see 
Eq.(A.5b)). We do not need the explicit form of ${Y}_{2r-1}^{(2q)}$ to 
prove that the right-hand side of Eq.(A.1) vanish.

Now let us write the expression for $1-R$ in the form of a series in 
powers of $kb$ with ${\nu }_{r}$ ($r \ge 1$) being the coefficients of 
this series. It is easy to see that since the coefficients
${Y}_{2r-1}^{(0)}$ are real numbers, ${\nu }_{1} = 0$. If $r > 1$, we 
have
$$
{\nu}_{r} = -i\frac{\mu}{\pi}\sum_{l=1}^{\infty}
\frac{{Y}_{2l-1}^{(r-1)}- {({Y}_{2l-1}^{(r-1)})}^{*}}{{(2l-1)}^{2}}
-{\left(\frac{\mu}{\pi}\right)}^{2}\sum_{l,m=1}^{\infty}
\frac{1}{{(2l-1)}^{2}{(2m-1)}^{2}}\sum_{q=0}^{r-2}
{Y}_{2m-1}^{(r-q-2)}{({Y}_{2l-1}^{(q)})}^{*}.  \eqno (A.9)
$$

Let us show that ${\nu}_{r}=0$ for an arbitrary $r$. Indeed, if $r=2s-
1$, taking into account that all the coefficients ${Y}_{2l-1}^{(2(s-
1))}$ are real numbers, we see that the first term in the right-hand 
side of Eq.(A.9) is equal to zero. In the second term decomposing the 
sum over $q$ into two sums over odd and even values of $q$, after a 
simple transformation we obtain
$$
{\nu}_{2s-1} = -{\left(\frac{\mu}{\pi}\right)}^{2}\sum_{l,m=1}^{\infty}
\frac{1}{{(2l-1)}^{2}{(2m-1)}^{2}}\sum_{q=0}^{s-2}{Y}_{2l-1}^{(2q)}
[{Y}_{2m-1}^{(2(s-q)-3)}+ {({Y}_{2m-1}^{(2(s-q)-3})}^{*}].  \eqno (A.10)
$$
and, consequently, ${\nu}_{2s-1}=0$.

Next, let us examine the case $r=2s$. If we take into account that all 
the coefficients ${Y}_{2l-1}^{(2s-1)}$ are imaginary numbers, from 
Eq.(A.9) we obtain
$$
{\nu}_{2r} = -2i\frac{\mu}{\pi}\sum_{l=1}^{\infty}
\frac{{Y}_{2l-1}^{(2r-1)}}{{(2l-1)}^{2}}
-{\left(\frac{\mu}{\pi}\right)}^{2}\sum_{l,m=1}^{\infty}
\frac{1}{{(2l-1)}^{2}{(2m-1)}^{2}}\sum_{q=0}^{2(r-1)}
{Y}_{2m-1}^{(2(r-1)-q)}{({Y}_{2l-1}^{(q)})}^{*}.  \eqno (A.11a)
$$
Now we make use of Eq.(A.9). From this equation it follows that
$$
\sum_{l=1}^{\infty}\frac{{Y}_{2l-1}^{(2s-1)}}{{(2l-1)}^{2}}=
\frac{i\mu}{2\pi}\sum_{l,m=1}^{\infty}
\frac{1}{{(2l-1)}^{2}{(2m-1)}^{2}}\sum_{q=0}^{2(s-1)}
{Y}_{2m-1}^{(q)}{({Y}_{2l-1}^{(2(s-1)-q)})}^{*}.  \eqno (A.11b)
$$
From the last equation we immediately have ${\nu}_{2s}=0$.

Thus we have shown that if $\zeta = 0$ and $kb < 1$, the solution of the 
matrix equation (18.1) provides the fulfillment of the equality $1-R=0$ 
in agreement with the energy conservation law.

\newpage

\centerline{\bf LIST OF FIGURES}

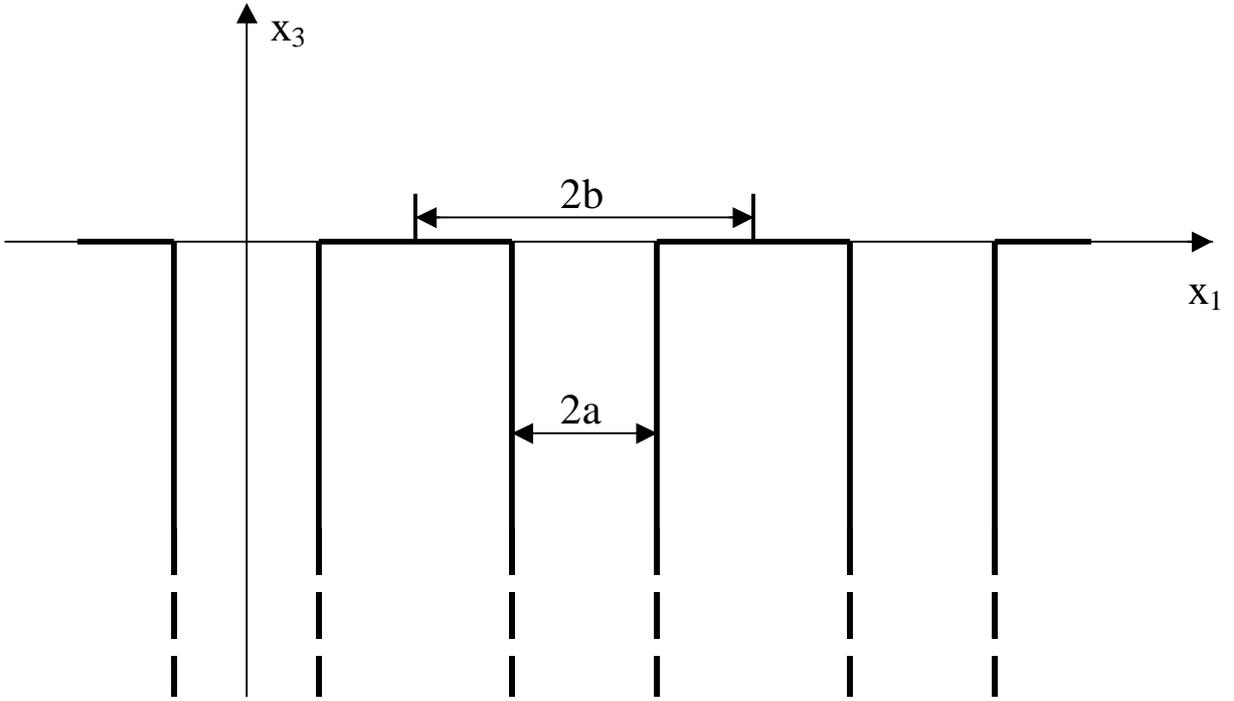
Fig.1. The grating configuration; $2b$ is the period, $2a$ is the width 
of the grooves throats.

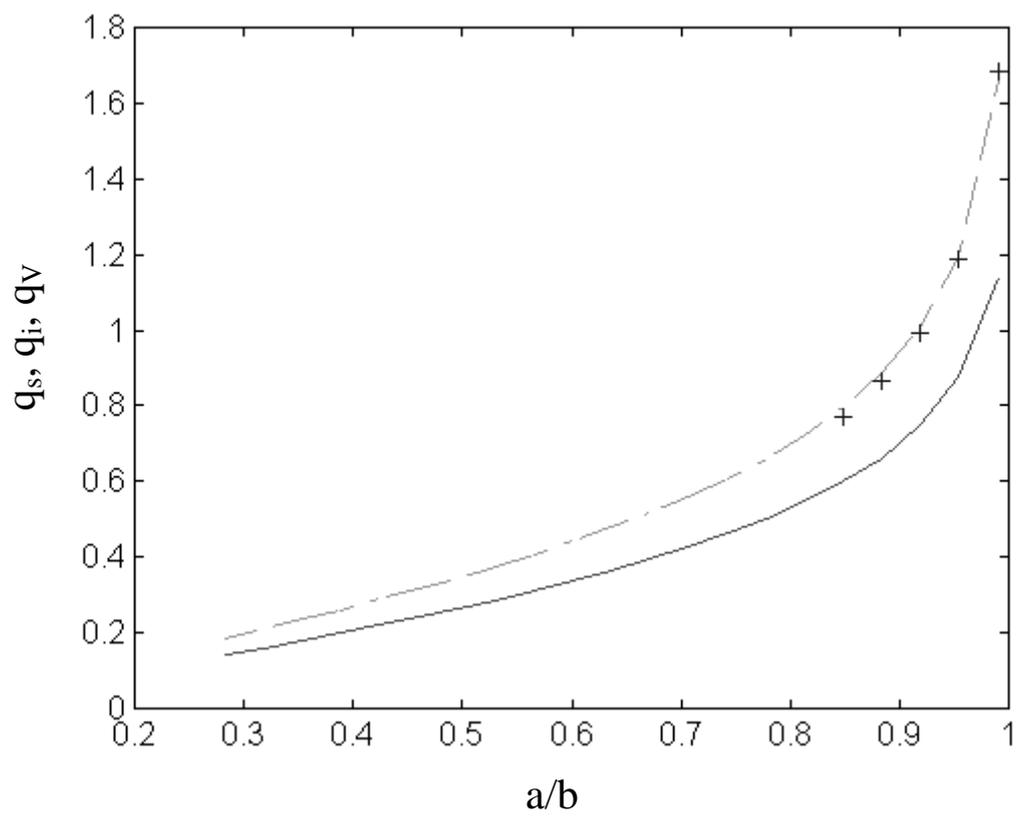
Fig.2. The dimensionless values $q_s = \pi Q_d^{(sur)}/cb{|E_0|}^{2}$ (solid 
line) and $q_i = \pi Q_d^{(\infty)}/cb{|E_0|}^{2}$ (dashed line) as functions 
of the ratio $a/b$. The crosses show the values of $q_V = \pi 
Q_d^{(V)}/cb{|E_0|}^{2}$ calculated with the aid of Eq.(26).

Fig.3. The ratios $N_R = {\rm Re}{\varphi}_{2n-1}(0)/ {\varphi}_{2n-
1}^{(0)}(0)$ (solid lines) and $N_I = {\rm Im}{\varphi}_{2n-1}(0)/ 
{\varphi}_{2n-1}^{(0)}(0)$ (dashed line) in the point $x_1=0$ for two 
values of $\beta$: $\beta = 0.05$ (Fig.3a) and $\beta = 0.01$ (Fig.3b).
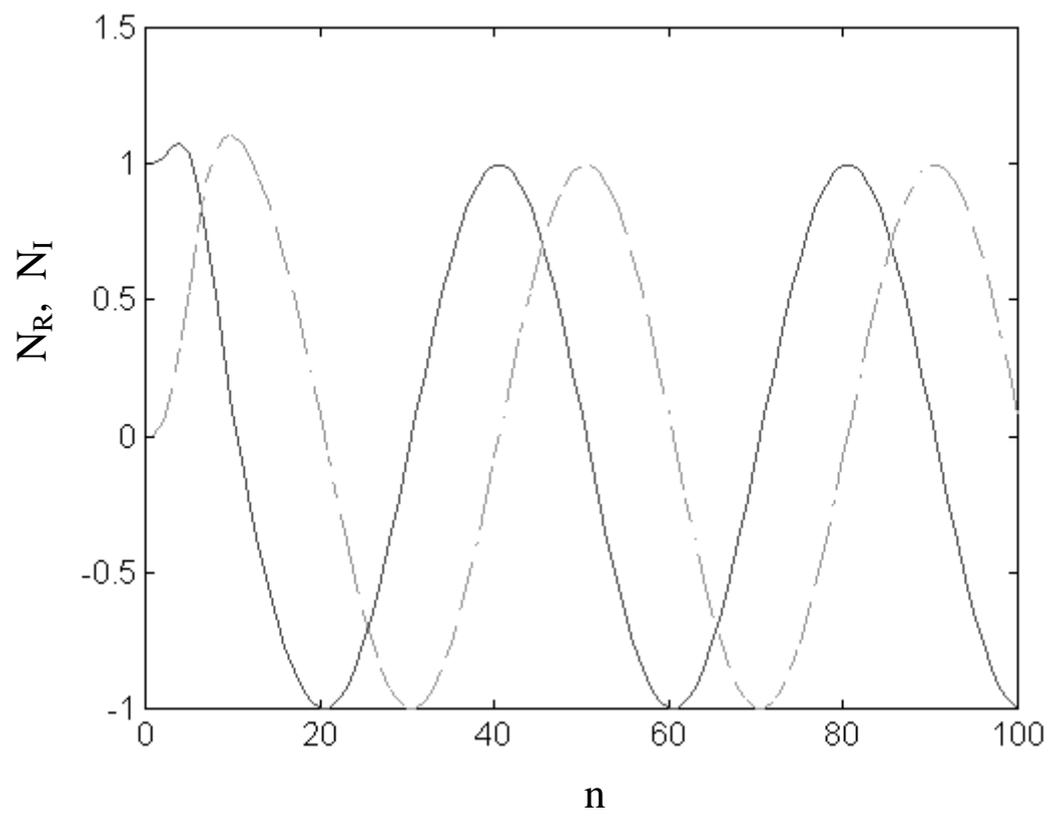
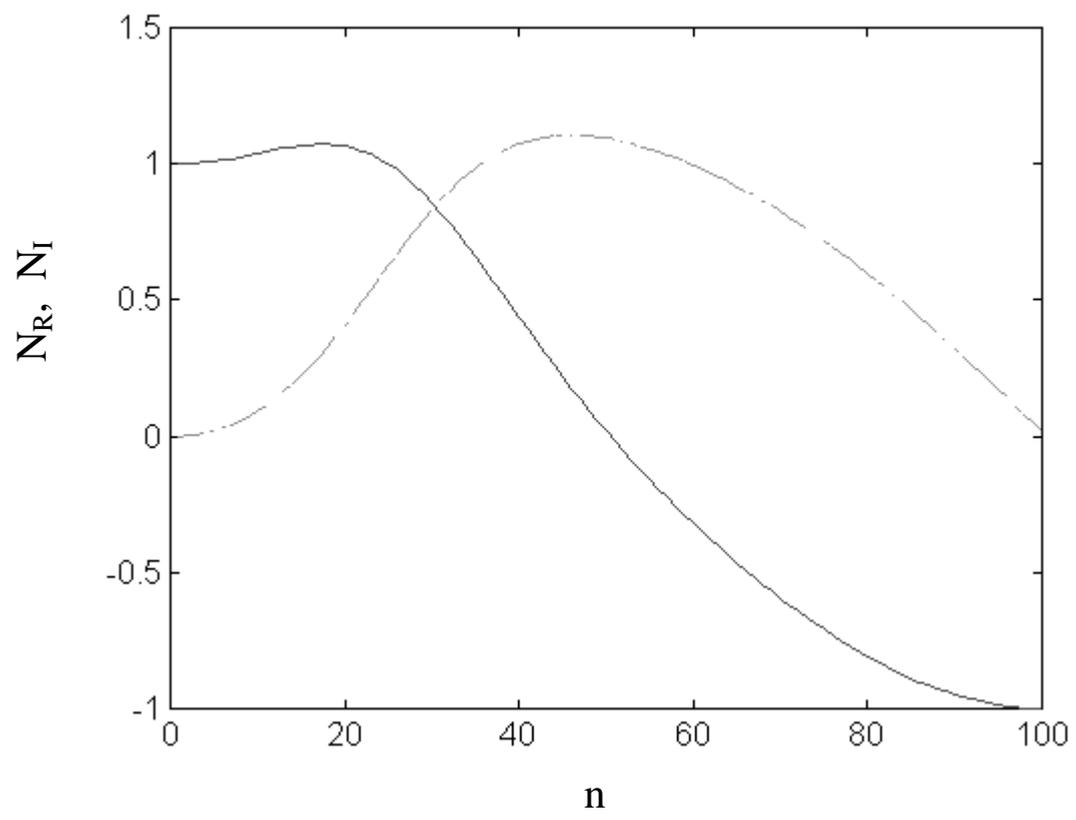
\end{document}